\documentclass[twocolumn,showpacs,preprintnumbers,amsmath,amssymb,floatfix]{revtex4}

\usepackage{epsf}

 \newcommand{\insertplot}[5]{\begin{figure}
 \hfill\hbox to 0.05in{\vbox to #5in{\vfill
 \inputplot{#1}{#4}{#5}}\hfill}
 \hfill\vspace{-.1in}
 \caption{#2}\label{#3}
 \end{figure}}
 \newcommand{\inputplot}[3]{
 \special{ps: plotfile #1}
}

\newcommand{\rd}{{\rm{d}}}

\begin{document}

\title{
$D=5$ Einstein-Maxwell-Chern-Simons Black Holes}
 \vspace{1.5truecm}
\author{
{\bf Jutta Kunz and  Francisco Navarro-L\'erida}
}
\affiliation{
{
Institut f\"ur Physik, Universit\"at Oldenburg, Postfach 2503\\
D-26111 Oldenburg, Germany}
}
\date{\today}
\pacs{04.20.Jb, 04.40.Nr}

\begin{abstract}
5-dimensional Einstein-Maxwell-Chern-Simons theory
with Chern-Simons coefficient $\lambda=1$
has supersymmetric black holes with vanishing horizon angular velocity,
but finite angular momentum.
Here supersymmetry is associated with a
borderline between stability and instability,
since for $\lambda>1$ a rotational instability arises,
where counterrotating black holes appear,
whose horizon rotates in the opposite sense to the angular momentum.
For $\lambda>2$ black holes are no longer uniquely characterized
by their global charges,
and rotating black holes with vanishing angular momentum appear.
\end{abstract}

\maketitle

{\sl Introduction}

Higher dimensional black holes received much interest in recent years,
in particular in the context of string theory,
and with the advent of brane-world theories, 
raising the possibility of direct observation 
in future high energy colliders \cite{exp}.

Static charged asymptotically flat black hole solutions of 
Einstein-Maxwell (EM) theory exist for all spacetime dimensions $D \ge 4$
\cite{tangher,MP}.
The generalization of the Kerr metric to higher dimensions
was obtained by Myers and Perry \cite{MP},
while the higher dimensional generalization of the Kerr-Newman
metric is still not known analytically.
So far rotating charged EM black holes
have only been found numerically \cite{KNP}.

EM black holes are characterized by their mass $M$, charge $Q$, and 
$[(D-1)/2]$ angular momenta $\bf J$, their number corresponding to the
rank of the rotation group SO(D-1) \cite{MP}.
Their event horizon has surface gravity $\kappa$ and $(D-2)$-volume
${\cal A}$, electrostatic potential $\Phi_{\rm H}$ and 
$[(D-1)/2]$ angular velocities ${\bf \Omega}_{\rm H}$.
These black holes satisfy the Smarr formula \cite{GMT}
\begin{equation}
M=  \frac{(D-2)}{(D-3)8\pi G_D} \kappa {\cal A}
+ \Phi_{\rm H} Q + \frac{(D-2)}{(D-3)}{\bf \Omega}_{\rm H} \cdot {\bf J} \ ,
\label{smarr}
\end{equation}
where $G_D$ is the $D$-dimensional Newton constant,
and the first law of black hole mechanics \cite{GMT}
\begin{equation}
\rd M= \frac{\kappa}{8\pi G_D} \rd {\cal A} + \Phi_{\rm H} \rd Q +
{\bf \Omega}_{\rm H} \cdot \rd {\bf J}\, .
\label{first}
\end{equation}

In odd dimensions $D=2n+1$ the Einstein-Maxwell action 
may be supplemented by a `$A\,F^n$' Chern-Simons (CS) term.
While not affecting the static black hole solutions,
this term does affect the stationary black hole solutions,
yielding a modified Smarr formula,
supplemented by an additional term
proportional to the CS coefficient $\lambda$ and
to $(D-5)$ \cite{GMT}, i.e., $D=5$ is a rather special case
among the class of odd-dimensional 
Einstein-Maxwell-Chern-Simons (EMCS) theories,
since the Smarr formula (\ref{smarr}) remains unmodified.

The bosonic sector of minimal $D=5$ supergravity
may be viewed as the special $\lambda=1$ case
of the general EMCS theory with Lagrangian
\begin{equation}
{\cal L}= \frac{1}{16\pi G_5} \left[\sqrt{-g}(R -F^2) -
\frac{2\lambda}{3\sqrt{3}}\varepsilon^{mnpqr}A_mF_{np}F_{qr}\right] \ ,
\label{Lag}
\end{equation}
and CS coefficient $\lambda$.
Surprisingly, the addition of the CS term 
makes it easier to solve the field equations
in the special case of the supergravity coefficient $\lambda=1$ \cite{Horo},
and analytic solutions describing charged, rotating black holes
are known \cite{BLMPSV,BMPV,Cvetic}.

The extremal limits of the $D=5$ rotating charged black hole
solutions of (\ref{Lag}) with $\lambda=1$
are of special interest, since they encompass a
two parameter family of stationary supersymmetric black holes \cite{BMPV}.
The mass of these supersymmetric black holes
is determined in terms of their charge
and saturates the bound \cite{Gibbons}
\begin{equation}
M \ge \frac{\sqrt{3}}{2} |Q| \ ,
\label{M-bound}
\end{equation}
and their two equal-magnitude angular momenta, $|J|=|J_1|=|J_2|$,
are finite and satisfy the bound \cite{BMPV,surprise}
\begin{equation}
|J|^2 \le \frac{1}{6 \sqrt{3} \pi}  |Q|^3 \ ,
\label{J-bound}
\end{equation}
in units for which $G_5 =1$.
However, their horizon angular velocities $\bf \Omega_{\rm H}$ vanish.
Thus their horizon is non-rotating, although their angular momentum
is nonzero. 
Clearly, angular momentum is stored in the Maxwell field,
but surprisingly, a negative fraction of the total angular momentum
is stored behind the horizon \cite{GMT,surprise}.
The effect of rotation on the horizon is not to make it rotate
but to deform it into a squashed 3-sphere \cite{GMT}.

These special properties of $D=5$ supersymmetric EMCS black holes
have caused intriguing speculations on how the properties
of $D=5$ black holes in general EMCS theories might depend
on the CS coefficient \cite{GMT}.
Centered on the issue of stability
these speculations involve an increase of the CS coefficient beyond its
supergravity value, i.e., beyond $\lambda = 1$.

The bound (\ref{M-bound}) on the mass of
$D=5$ EMCS black holes relies on the fact
that $\lambda=1$ \cite{Gibbons,GMT}.
But extremal static EMCS black holes 
saturate the bound for any value of $\lambda$.
If the mass of extremal stationary black holes 
decreases with increasing $\lambda$ for fixed angular momentum,
and increases with increasing angular momentum for fixed $\lambda<1$,
while it is independent of angular momentum for $\lambda=1$,
it becomes conceivable that the mass can
decrease with increasing angular momentum for fixed $\lambda>1$ \cite{GMT}.
Thus while an extremal static black hole 
with zero Hawking temperature and spherical symmetry 
cannot decrease its mass by Hawking radiation,
it could however become unstable with respect to rotation, when $\lambda>1$,
with photons carrying away both energy and angular momentum
to infinity \cite{GMT}.
In terms of the first law as applied to $\lambda>1$
extremal black holes ($\kappa=0$) with fixed charge ($\rd Q=0$),
such an instability would require, that 
the horizon would be rotating in the opposite sense to the angular momentum,
since $\rd M = {\bf \Omega}_{\rm H} \cdot \rd {\bf J}$
would have to be negative \cite{GMT}.
In such a case supersymmetry would once again be associated with a
borderline between stability and instability \cite{GMT}.

By numerically constructing charged rotating $D=5$ EMCS black holes
for arbitrary values of $\lambda$, we here verify these speculations,
and address intriguing similarities with 
$D=4$ Einstein-Maxwell-dilaton (EMD) black holes 
\cite{Gibbons,Rasheed,KKN-c}.
We restrict to black holes with horizon topology of a sphere
\cite{blackrings},
and show that for $\lambda>2$, these black holes are
no longer uniquely characterized by their global charges.

{\sl EMCS black holes}

To obtain stationary EMCS black hole solutions,
representing charged generalizations of the 5-dimensional 
Myers-Perry solutions \cite{MP},
we consider black hole space-times with bi-azimuthal symmetry,
implying the existence of three commuting Killing vectors,
$\xi = \partial_t$, $\eta_1=\partial_{\varphi_1}$, 
and $\eta_2=\partial_{\varphi_2}$
\cite{MP,frolov}.
We employ a parametrization for the metric 
based on bi-azimuthal isotropic coordinates, 
well suited for numerical work \cite{KNP,kkrot}
\begin{eqnarray}
&&\rd s^2 = -f \rd t^2
  + \frac{l_1}{f}\, r^2 \sin^2\theta
          \left( \rd \varphi_1-\frac{\omega_1}{r} \rd t\right)^2
\nonumber \\
&&+ \frac{l_2}{f}\, r^2 \cos^2\theta
          \left( \rd \varphi_2-\frac{\omega_2}{r} \rd t\right)^2
 +\frac{m}{f}\left( \rd r^2+r^2 \rd \theta^2 \right)
 \nonumber \\ &&
 +\frac{p}{f}\, r^6 \sin^2\theta \cos^2\theta  
 \left( \omega_2 \rd \varphi_1 - \omega_1 \rd \varphi_2 \right)^2 
          \ . \label{metric} \end{eqnarray}

The metric has two orthogonal 2-planes of rotation, 
corresponding to $\theta=0$ and $\theta=\pi/2$.
The gauge potential is parametrized by \cite{frolov}
\begin{equation}
A_{\mu} \rd x^{\mu}=A_t \rd t
+A_{\varphi_1} \rd \varphi_1+A_{\varphi_2} \rd \varphi_2 \ . 
\end{equation}
All metric and gauge field functions depend on $r$ and $\theta$ only.

At infinity we impose on the metric the boundary conditions
$f=m=l_i=1$, $p=\omega_i=0$, 
i.e., the solutions are asymptotically flat,
and on the gauge field $A_t=A_{\varphi_i}=0$.

The regular event horizon resides at a surface of constant radial coordinate,
$r=r_{\rm H}$ \cite{MP,KNP},
and is characterized by the condition $f(r_{\rm H})=0$ \cite{kkrot,KKN-c}.
Here the metric functions satisfy the boundary conditions
$f=m=l_i=p=0$, $\omega_i=\omega_{i,\rm H}$,
where $\omega_{i,\rm H}$ are constants
determining the horizon angular velocities
$\Omega_i=\omega_{i,\rm H}/r_{\rm H}$,
and the gauge field satisfies
$\left. \chi^\mu A_\mu \right|_{r_{\rm H}} = - \Phi_{\rm H}$, 
$\partial_r A_{\varphi_i}=0$,
with Killing vector
$\chi=\xi + \Omega_1 \eta_1 + \Omega_2 \eta_2$.

The boundary conditions in the planes $\theta = 0$ and $\theta = \pi /2$
are determined by symmetries. In both planes
$\partial_{\theta}f=\partial_{\theta}m=
\partial_{\theta}l_i=
\partial_{\theta}p=
\partial_{\theta}\omega_i= \partial_{\theta} A_t =0$,
and in the $\theta = 0$ plane
$\partial_\theta A_{\varphi_2}=0$, $A_{\varphi_1}=0$,
while in the $\theta = \pi /2$ plane
$\partial_\theta A_{\varphi_1}=0$, $A_{\varphi_2}=0$.

The mass $M$ and the angular momenta $J_i$ of the black hole 
are obtained from the asymptotic expansion for the metric 
\begin{equation}
f \rightarrow 1-\frac{8\, G_5 M }{3\pi r^2} \ , \ \ \
\omega_i \rightarrow \frac{4\, G_5 J_i }{\pi r^3} \ ,
\end{equation}
while the charge $Q$ and the magnetic moments ${\cal M}_i$ 
are obtained from the asymptotic expansion for the gauge potential 
\begin{equation}
A_t \rightarrow - \frac{G_5 Q}{\pi r^2}  \ , \ \ \
A_{\varphi_i} \rightarrow  \frac{G_5 {\cal M}_i \sin^2\theta}{ \pi r^2} 
\ . \end{equation}

\begin{figure}[t]
\begin{center}
\epsfysize=6.5cm
\mbox{\epsffile{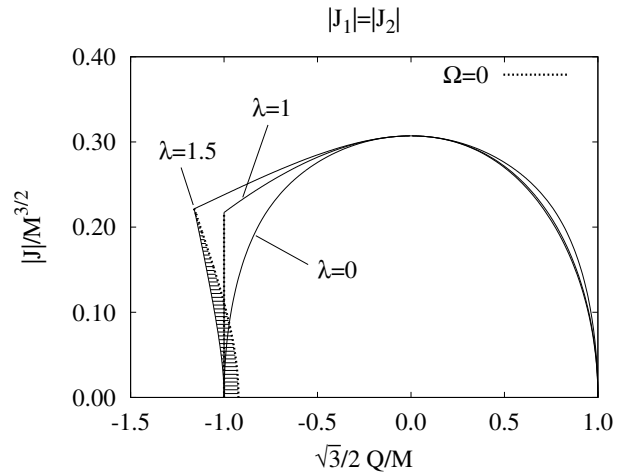}}
\caption{
Scaled angular momentum $|J|/M^{3/2}$ versus
scaled charge $Q/M$ for extremal black holes (solid)
and stationary black holes with non-rotating horizon (dotted)
($\lambda=0$, 1, 1.5).
}
\end{center}
\end{figure}

The expansion at the horizon shows, that the surface gravity $\kappa$
is constant at the horizon,
as required by the zeroth law of black hole mechanics,
and that the electrostatic potential $\Phi_{\rm H}$
is constant at the horizon as well.
To have a measure for the deformation of the horizon we consider
the circumferences of the horizon,
$L_1$, where $\varphi_1=\varphi_2=$ const,
$L_2$, where $\varphi_2=\theta=$ const,
and $L_3$, where $\varphi_1=\theta=$ const.


{\sl Numerical results}

To perform the numerical calculations, we introduce
the compactified radial variable $\bar{r}= 1-r_{\rm H}/r$ \cite{kkrot,KNP}.
We focus on black holes with two equal-magnitude angular
momenta, $|J|=|J_1|=|J_2|$. 
Then the angular dependence of the functions can be extracted,
and a system of 6 ordinary differential equations remains to be solved \cite{long}.


In Fig.~1 we exhibit
the scaled angular momentum $|J|/M^{3/2}$ of extremal EMCS black holes 
versus the scaled charge $Q/M$ \cite{foot1}
for three values of $\lambda$:
the pure EM case, $\lambda=0$ \cite{KNP}, 
the supergravity case, $\lambda=1$ \cite{BMPV}, 
and for $\lambda=1.5$, a value beyond the supergravity value.
For a given value of $\lambda$ 
black holes exist only in the regions bounded by the
$J=0$-axis and by the respective solid curves.
Note the asymmetry of the domain of existence of the black hole solutions
with respect to $Q \rightarrow -Q$
for non-vanishing CS term \cite{foot2}.

\begin{figure}[t]
\begin{center}
\epsfysize=6.5cm
\mbox{\epsffile{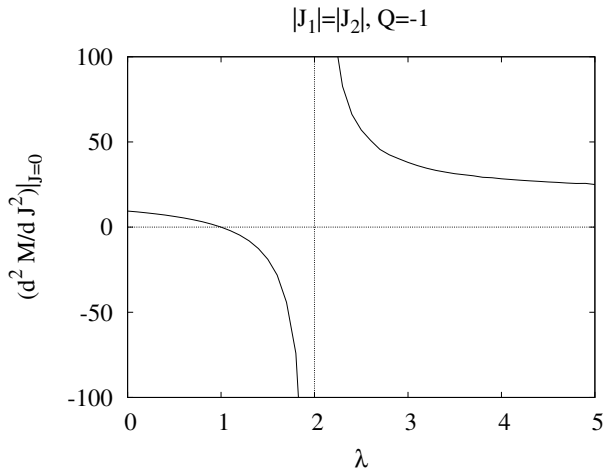}}
\caption{
Second-order derivative of the mass $M$ with respect to the angular momentum
$J$ at $J=0$ versus the CS coupling constant $\lambda$ for extremal
black holes ($Q=-1$).
}
\end{center}
\end{figure}

The dotted curves in the figure
correspond to stationary black holes with non-rotating horizon, i.e.,
to black holes with horizon angular velocity $\Omega=0$.
Such solutions are present only for $\lambda \ge 1$.
For $\lambda=1$ they are extremal solutions,
forming the vertical part of the $Q<0$ borderline.
For $\lambda>1$, however, they are non-extremal black holes,
and thus located within the allowed region,
with the exception of a single point, which is part of the borderline.
In fact, these $\Omega=0$ black holes divide this region into two parts.
The right part contains ordinary black holes,
where the horizon rotates in the same sense as the angular momentum.
But the left part (the shaded region) contains extraordinary black holes.
Their horizon rotates in the opposite sense to the angular momentum,
therefore we refer to them as counterrotating black holes \cite{KKN-c}.
Counterrotating black holes appear only for $\lambda > 1$. The bound (\ref{M-bound}) is clearly violated, when $\lambda > 1$.

\begin{figure}[t]
\begin{center}
\epsfysize=6.5cm
\mbox{\epsffile{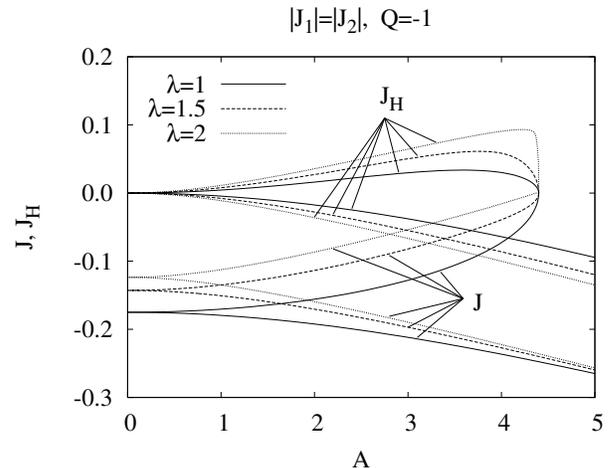}}
\caption{
Total angular momentum $J$ and 
horizon angular momentum $J_{\rm H}$
versus area ${\cal A}$
for (almost) extremal black holes
($\lambda=1$, 1.5, 2; $Q=-1$).
}
\end{center}
\end{figure}

In Fig.~2 we demonstrate explicitly,
that extremal static black holes can become
unstable with respect to rotation,
by exhibiting the deviation of the mass from the
static value, computed via the second-order derivative of the
mass with respect to the angular momentum at the extremal static solution.
Clearly, for $1< \lambda < 2$ the mass is seen to decrease
with increasing magnitude of the angular momentum for fixed electric charge,
because of the presence of counterrotating solutions,
connected to the extremal static solution.
Thus we observe indeed, 
that supersymmetry marks again a
borderline between stability and instability.

In Fig.~3 we exhibit the total angular momentum $J$ 
and the horizon angular momentum $J_{\rm H}$ \cite{GMT}
as functions of the area ${\cal A}$
for (almost) extremal black holes with $Q=-1$ and $\lambda=1$,
$\lambda=1.5$, and $\lambda=2$.
When $\lambda=1$,
starting from a static black hole with finite area, the supersymmetric
black holes decrease in size with increasing $|J|$, saturating the
bound (\ref{J-bound}) in the limit ${\cal A}=0$.
At the same time the black holes become more and more squashed,
as seen from their ratio of circumferences.
For these supersymmetric black holes
$J$ and $J_{\rm H}$ have opposite signs, while $\Omega=0$.
Since $\rd M = \Omega \, \rd J =0$,
their mass $M$ remains constant, 
while the angular momentum is built up.
Along the non-supersymmetric branch $J$ and $J_{\rm H}$ have equal signs.
As $\lambda$ is increased, the (continuous)
set of $\Omega=0$ solutions becomes non-extremal, 
whereas below $\lambda=1$ the only $\Omega=0$ solutions are static.

As expected from the change in stability,
another special case is reached, when $\lambda=2$.
Indeed, as $\lambda$ is increased beyond $2$, another new phenomenon arises:
a (continuous) set of non-extremal rotating $J=0$ solutions appears 
\cite{strau}.
Their existence relies on a special partition of the total
angular momentum $J$, where the angular momentum within the horizon 
$J_{\rm H}$ is equal and opposite to 
the angular momentum in the Maxwell field outside the horizon \cite{foot3}.
In contrast, for $\lambda<2$ only static $J=0$ solutions exist.
The presence of $J=0$ solutions is exhibited in Fig.~4 for $\lambda=3$.

Related to the $J=0$ solutions, the domain of existence of EMCS black holes 
changes, and an extremal rotating $J=0$ solution replaces
the extremal static solution as the left boundary point on the $J=0$ axis,
when $\lambda>2$. 
As $\lambda$ is increased further, more
branches of $J=0$ solutions and
$\Omega=0$ solutions appear \cite{long}.

The numerical data indicate, that at $\lambda=2$
a (continuous) set of extremal rotating $J=0$ black holes with constant mass 
is present.
As the horizon angular velocity $\Omega$ increases,
their mass $M$ can remain constant, as long as $J=0$,
and the angular momentum is redistributed appropriately
(as indicated by the steep rise of $J_{\rm H}$ in Fig.~3).
For these black holes, the deformation is oblate.

\begin{figure}[t]
\begin{center}
\epsfysize=6.5cm
\mbox{\epsffile{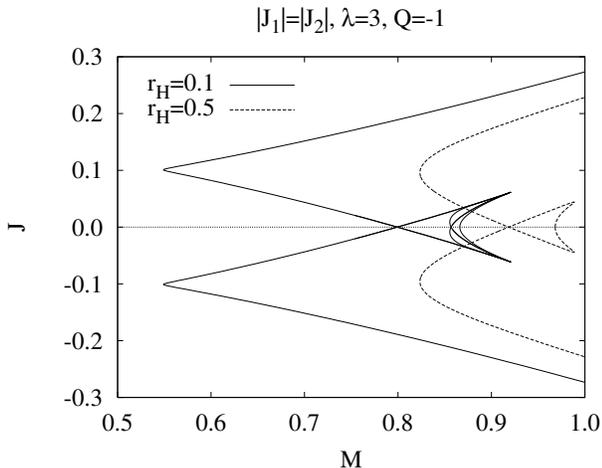}}
\caption{
Angular momentum $J$ versus mass $M$
for non-extremal black holes with horizon radii $r_{\rm H}=0.1$ and 0.5
($\lambda=3$; $Q=-1$).
}
\end{center}
\end{figure}

Fig.~4 further reveals that beyond $\lambda=2$ 
black holes are no longer uniquely characterized by their global charges.
Thus the uniqueness conjecture does not hold for $D=5$ EMCS
stationary black holes with horizons of spherical topology,
provided $\lambda>2$ \cite{foot4}.
The previously known counterexamples involved black rings \cite{blackrings}.

{\sl Comparison with $D=4$ EMD black holes}

A number of features of 
$D=5$ EMCS black holes appear also for $D=4$ EMD black holes,
with dilaton coupling constant $\gamma$,
when both electric ($Q$) and magnetic ($P$) charge are present \cite{KKN-c}.
In EMD theory the Kaluza-Klein value $\gamma=\sqrt{3}$
represents the critical value.
For $\gamma<\sqrt{3}$ only corotating black holes exist.
For $\gamma=\sqrt{3}$ stationary black holes with non-rotating horizon
appear and form the vertical part of the boundary \cite{Rasheed}.
Their angular momentum is bounded by $|J| \le |PQ|$ 
(analogous to (\ref{J-bound})).
For $\gamma>\sqrt{3}$ corotating and counterrotating black holes 
exist \cite{KKN-c}.
Thus in EMD theory the Kaluza-Klein value $\gamma=\sqrt{3}$ marks 
the change from stability to instability.
Stationary $\Omega=0$ black holes and counterrotating
black holes also exhibit squashed horizons \cite{KKN-c}.

This analogy raises the question,
in which theories under which circumstances
such features of stationary black holes arise.
A hint may be, that for EMCS and EMD theory, 
the respective critical values of the coupling constants
each yield a theory with a high degree of symmetry.

%


\begin{thebibliography}{000}
\bibitem{foot1}
 The solutions have a scaling symmetry
 $\tilde M= \mu^2 M$,
 $\tilde J= \mu^3 J$,
 $\tilde Q = \mu^2 Q$,
 $\tilde r_{\rm H}=\mu r_{\rm H}$,
 $\tilde \Omega = \Omega/\mu$, 
 $\tilde \kappa = \kappa/\mu$, etc.
\end{thebibliography}
\end{document}